# New Globular Cluster Age Estimates and Constraints on the Cosmic Equation of State and The Matter Density of the Universe.

Lawrence M. Krauss* & Brian Chaboyer

*Departments of Physics and Astronomy, Case Western Reserve University, 10900 Euclid Ave., Cleveland OH USA 44106-7079;   Department of Physics and Astronomy, Dartmouth College, 6127 Wilder Laboratory, Hanover, NH. USA

**New estimates of globular cluster distances, combined with revised ranges for input parameters in stellar evolution codes and recent estimates of the earliest redshift of cluster formation allow us to derive a new 95% confidence level lower limit on the age of the Universe of 11 Gyr.  This is now definitively inconsistent with the expansion age for a flat Universe for the currently allowed range of the Hubble constant unless the cosmic equation of state is dominated by a component that violates the strong energy condition.  This solidifies the case for a dark energy-dominated universe, complementing supernova data and direct measurements of the geometry and matter density in the Universe. The best-fit age is consistent with a cosmological constant-dominated (w=pressure/energy density = -1) universe.  For the Hubble Key project best fit value of the Hubble Constant our age limits yields the constraints w < -0.4 and $\Omega_{matter}$ < 0.38  at the 68 % confidence level, and w < -0.26 and $\Omega_{matter}$ < 0.58 at the 95 % confidence level.**

Age determinations of globular clusters provided one of the earliest motivations for considering the possible existence of a cosmological constant.  By comparing a lower limit on the age of the oldest globular clusters in our galaxy--- estimated in the 1980 s to be 15-20 Gyr---with the expansion age, determined by measurements of the Hubble constant, an apparent inconsistency arose:  globular clusters appeared to be older than the Universe unless one allowed for a possible Cosmological Constant. Introduction of such a term results in an increased expansion age for a given Hubble constant because it implies that the expansion rate at earlier times was slower than would otherwise be the case, implying that galaxies located a certain distance away from us today would have taken longer to achieve this separation since the Big Bang.

Unfortunately this constraint did not turn out to be robust. As more careful examinations of uncertainties associated with stellar evolution were performed, and well as refined estimates of the parameters that govern stellar evolution itself, the lower limit on globular cluster ages got progressively lower and lower, so that a wide range of cosmological parameters produced Hubble ages consistent with this lower limit[1,2,3]

As part of an ongoing program to refine globular cluster age estimates we are now able to derive a more stringent lower bound on globular cluster ages, due in part to a reanalysis of RR-Lyrae distance estimates but also in part to new estimates of the primordial helium abundance and a re-examination of the uncertainties associated with measurements of oxygen in low metallicity clusters. In addition, recent measurements of the Lyman Alpha forest, combined with new observations of large scale structure allow one to put a more stringent lower bound on the earliest epoch of globular cluster formation. When we combine these two new estimates, we derive a 95% lower bound on the age of the Universe of 11.2 Gyr. At the same time, CMB determinations of the curvature of the Universe suggest we live in a flat universe[4]. When we combine this result with our new age limits, some form of dark energy is required by the data. The best fit value for the age of the the Universe picks out a parameter range which is the same as that suggested by other, independent estimates of the dark matter and dark energy densities, while the edge of the range provides new upper limits on the dark matter density which are competitive with other independent estimators. We believe these new age estimates provide an important independent confirmation that dark energy dominates the energy density of the Universe. It is significant that the three fundamental probes of cosmological parameters: the Hubble expansion, the determination of the matter density and geometry of the Universe, and the determination of the age of the Universe, now separately and independently point to the same cosmological model to describe our Universe on large scales.

**Main Sequence Turn-off Age Estimates: An Overview of the Monte Carlo Method.**

Numerical stellar evolution models provide a general understanding of the preferred location of stars on a temperature-luminosity plot. In principle this agreement between theory and observation can be used to derive an age for a globular cluster. As stars evolve, their location on a temperature luminosity plot evolves. Thus the distribution on this plot for system of stars of different masses matches the theoretical distribution at only one single time. While probing the entire distribution would, in principle, provide the best constraint on the age of the system, in practice stellar models are best determined for main sequence stars. Thus, the main sequence turnoff luminosity, which occurs as a main sequence star exhausts its supply of hydrogen fuel, and which is a sensitive function of stellar ages, is used[1].

The theoretical stellar isochrones are dependent on a variety of parameters that are used as inputs for the numerical codes used to model evolving stars. These input parameters are subject to various theoretical and experimental uncertainties. In addition, in order to match the theoretical temperature-luminosity plots with observed stellar distributions, one must translate apparent magnitudes into luminosities, and colors into temperatures. Both of these translations are also subject to various uncertainties.

We have previously utilized Monte Carlo methods to evolve over 4 million stars, with input parameters chosen from distributions fit to the inferred uncertainties in input parameters, in order to derive a distribution of stellar isochrones to be compared to observed stellar distributions for low metallicity globular clusters. In comparing these distributions, we took into account observational uncertainties as well, allowing us to derive an age distribution for the oldest globular clusters with self-consistent 95% upper and lower confidence limits. We determined that the uncertainty in translating magnitudes to luminosities, i.e. the uncertainties in deriving distances to globular clusters, was the chief source of uncertainty in globular cluster age estimates, while at the same time we were able to catalogue and produce analytic dependences for globular cluster ages as a function of each of the stellar input parameters.

As new data arises on the key parameters affecting age estimates, we have re-examined the implications for putting a lower limit on the age of the Universe. This is particularly important for attempting to distinguish between different cosmological models. In particular, a matter dominated flat universe has a Hubble age that is 50% smaller than a flat Universe with a dark energy component making up 70% of the total energy density. Up until the present time, however, the residual uncertainties in globular cluster age estimates, combined with the uncertainty in the Hubble constant, implied that ages could not by themselves distinguish between these two possibilities.

**Globular Cluster Distance Estimates**

During the majority of its life, a star converts hydrogen into helium in its core. The most robust prediction of the theoretical models is the time it takes a star to exhaust its supply of hydrogen in the core[5]. Thus, absolute globular cluster age determinations with the smallest theoretical uncertainties are those that are based upon the luminosity of the main sequence turn-off. In order to determine the luminosity of a globular cluster turn-off star, however, one must determine the distance to the globular cluster.

There are a number of different methods that can be used to estimate the distance to globular clusters. All of these methods are secondary distance indicators, which rely upon various assumptions, or prior calibration steps.  While the distance scale to globular clusters is thus subject to many uncertainties, our knowledge of the distance scale is evolving rapidly.  Many of the distance indicators to globular clusters utilize the horizontal branch (HB) stars (stars which are converting helium into carbon in their cores) as standard candles.  RR Lyra stars are a subclass of HB stars which, due to instabilities in their atmospheres, are pulsating radially.  Our knowledge of the evolution of HB stars continues to advance through the use of increasingly more realistic stellar models. Recently, Demarque et al.[6]  have made a comprehensive study of the evolution of HB stars, and their use as distance indicators.  A key point made in their study is that the luminosity of the HB stars depends not only on metallicity, but also on the evolutionary status of the stars on the HB in a given globular cluster.  Other theoretical calculations have come to similar conclusions[7].

Earlier analyses have tended to assume that there was a simple linear relation between HB luminosity and the magnitude of the RR Lyra stars.  The work of Demarque et al.  requires us to take a new approach to determine the distance scale to globular clusters.  In particular, we have elected to concentrate our attention on those distance determination techniques that can best yield distances to metal-poor globular clusters, since these are the clusters of interest in deriving upper limits on the age of our galaxy. This alleviates much of the potential systematic error that can result when metal-rich stars are used to calibrate the distance to metal-poor clusters.

In order to inter-compare different distance indicators, it is convenient to parameterize the distance estimate by what it implies for the visual (V) magnitude of an RR Lyra star, $M_v(RR)$.  The results are summarized in Table 1 for the different distance indicators. There are three new features of this compilation, as compared to those associated with previous analyses: (1) the results of Gratton[8] 1998, who used Hipparocos parallaxes for metal-poor, blue HB stars in the field to calibrate the globular cluster distance scale is included; (2) the statistical parallax results on field RR Lyrae stars are included; (3) a new HST parallax for the star RR Lyrae itself which is considerably more accurate than the Hipparcos parallax[9]; and (4) only distance estimates for systems with [Fe/H] < -1.4 are included.

The mean [Fe/H] value of the objects used to determine the distance scale is shown in column 3, while column 4 gives the value of $M_v$(RR) at that value of [Fe/H]. In order to compare the different distance estimates, we need to translate these $M_v$(RR) values to a common [Fe/H] value. For this, we have elected to use an $M_v$(RR)--- [Fe/H] slope of 0.23– 0.06, as suggested by the theoretical models of Demarque et al.. We have chosen [Fe/H]= -1.9 for the common [Fe/H] value, as this is the mean [Fe/H] of the globular clusters whose average age we wish to determine. Since the different distance estimates span a relatively modest range in [Fe/H] (0.54 dex), the exact value of the $M_v$(RR)-- [Fe/H] has only a minor effect in our resultant distance scale. The weighted mean value of the absolute magnitude of the RR Lyrae stars at [Fe/H]= -1.9 is $M_v$(RR)= 0.47 mag.

As has been known for some time, the statistical parallax technique yields values for $M_v$(RR) which are a fair bit larger (ie. fainter) than the other distance techniques. When statistical parallax results are included in the weighted mean, the standard deviation about the mean is 0.13 mag. When the statistical parallax results are not included in the analysis, the mean becomes, $M_v$(RR)= 0.44 and the standard deviation about the mean drops to 0.07 mag. In earlier analyses the statistical parallax data was not included, as there were suggestions that some systematic differences might exists between RR Lyrae stars in the field, and in globular clusters. However, subsequent investigations have shown that this is not the case[10].

It is clear that if we are to include the statistical parallax in our analysis a determination of the allowed range in our adopted distance scale is somewhat subtler. If we use the – 0.13 mag standard deviation value that results from a naïve inclusion of the large value of $M_v(RR)$ from statistical parallax data in deriving the weighted mean, this results in a longer tail at *low* values of $M_v(RR)$ than would result if one did not incorporate the statistical parallax data in determining the weighted mean. It is clearly inappropriate to include this spurious low tail when quoting an allowed range. We have found that the asymmetric Gaussian distribution $M_v(RR) = 0.47^{+0.13}_{-0.10}$ mag has a low range is in agreement with that derived when the statistical parallax result is not included, but has mean and high range equivalent to that one would derive by including the statistical parallax result in a straightforward way. This is the distribution we thus use in deriving the allowed distance scale for metal-poor globular clusters.

**Stellar Evolution Input Parameters**

We have identified 7 critical parameters used in the computation of stellar evolution models whose estimated uncertainty can significantly affect derived globular cluster age estimates. In order of importance, they are (i) [O/Fe], (ii) mixing length, (iii) primordial helium abundance, (iv) $^{14}N + p \rightarrow {}^{15}O + \gamma$ reaction rate, (v) helium diffusion coefficient, (vi) color transformations, and (vii) low temperature opacities. The mean values and distributions for three of these parameters--- [O/Fe], the primordial helium abundance, and the helium diffusion coefficients---have changed substantially recently as the result of new measurements. We review these changes below.

Oxygen is the most abundant α-capture element, and by itself accounts for over 50% (by mass) of all elements heavier than helium in a star. As a result, the exact amount of oxygen that is used in the stellar evolution codes has a large impact on the derived globular cluster ages. In the last few years, several observational studies have appeared, which have determined the oxygen abundance in metal-poor stars[11,12,12,14,15].

These studies have found [O/Fe] in the range of 0.21 dex to 0.70 dex at [Fe/H] = -1.9. The estimated uncertainties given by individual groups are fairly small (of order – 0.06 dex), and yet different authors can analyze the same stars with quite different results. It is clear that the differences between the authors are systematic in nature. For this reason, we have elected to take a flat distribution for the oxygen abundance, with [O/Fe] in the range 0.2 to 0.7 dex. The mean of this range is [O/Fe] = 0.45 dex, which is 0.10 dex lower than previous estimates. This results in an increase in the derived globular cluster age estimate by +0.4 Gyr.

There have recently been considerable advances in our ability to estimate the primordial helium abundance. First, observations of deuterium in high redshift QSO absorption lines now allow a reliable estimation, using arguments from Big Bang Nucleosynthesis (BBN), of the cosmic baryon fraction. This value is $\Omega_B h^2 = 0.02 - .002$ (95%)[16] where the dependence on h represents our existing uncertainty in the Hubble constant, $H = 100 \, h \, \text{kms}^{-1}\text{Mpc}^{-1}$. This is in good agreement with the recently determined value $\Omega_B h^2 = 0.021 - .006$ (95%) from recent cosmic microwave background experiments, which yield a direct estimate for the baryon fraction in the universe by measuring the relative height of the first two peaks in the angular power spectrum[4]. The agreement between these two independent estimates is compelling and allows us to interpret the bound on the baryon fraction, using BBN, in terms of a new allowed range for the primordial helium abundance.

Based on the above arguments we assume a range for the helium to hydrogen mass fraction, Y, of 0.245-0.250, with a mean value of .2475. This is an increase in the mean helium abundance value by .0125 compared to earlier estimates, leading to a decrease in our best estimate of the age of the oldest globular clusters of 0.35 Gyr. Note that the choice of upper limit of .250 on Y is conservative in this regard in that it allows for a more conservative lower limit on globular cluster ages.

Finally, recent observations of iron in the turn-off stars of metal-poor globular clusters have lead to a revision in our understanding of the importance of helium diffusion in the evolution of metal-poor stars. This process occurs in stable plasmas, whereby helium sinks to the center of a star, displacing some hydrogen in the core and shortening the main sequence lifetime. Helioseismology has produced conclusive evidence that helium diffusion is occurring in the Sun[17,18]. However, Chaboyer et al[19] have shown that because the iron abundance in turn-off stars in a metal-poor globular cluster is identical to that in stars on the giant branch in the same globular cluster helium diffusion must be inhibited near the surface of the metal-poor stars. Exactly why this is the case is currently not known. Using stellar models in which diffusion was inhibited on the surface these authors found that the resultant isochrones were very similar to those in which the helium diffusion coefficients had been cut in half from their nominal value. Since a priori calculations of the helium diffusion coefficients are in any case uncertain at at least the 30% level[20], we have elected to multiply the nominal diffusion coefficients by a factor, D which is in the range D = 0.2 - 0.8. The midpoint of this range, D = 0.5 corresponds to the inhibited diffusion case described above. This should be compared to the mean value D = 0.75 used previously, and this change in has the effect of increasing derived ages by +0.2 Gyr.

**The age and formation time of Globular Clusters and Constraints on the Cosmic Equation of State and Mass Density**

Utilizing the estimates for the parameter ranges described in the preceding sections, along with standard parameter ranges for the other stellar evolution variables[1,2] we present in figure 1 the results of our Monte Carlo fit of predicted isochrones to observed isochrones to determine the age of the oldest galactic globular clusters. Taking a one-sided 95 % range we find a lower bound of 10.2 Gyr, a 1 Gyr increase compared earlier estimates. This increase is due to a confluence of factors, although it is dominated the new Mv-RR distance estimates. Note that the best fit age has also increased by an equivalent amount, and is now 12.4 Gyr.

In order to utilize these results to constrain cosmological parameters, we need to add to these ages a time which corresponds to the time between the Big Bang and the formation of globular clusters in our galaxy. An upper limit on this time is thought to be 2 Gyr. However, for our purposes, we must estimate a lower limit.

Several recent studies, utilizing observations of globular clusters in nearby galaxies, and measurements of the redshift of Lyman $\alpha$ systems, all independently appear to put a limit $z < 5$ for the maximum redshift of structure formation on the scale of globular clusters[21,22]. Nevertheless, to be conservative, we will assume $z < 6$ for our lower age estimates here.[1a]

In order to turn this redshift into a time, in principle one has to assume a cosmological model. Fortunately, however, the age of the universe as a function of redshift is largely insensitive to the cosmic equation of state today, for redshifts greater than about 3-4. This is because for these early times the matter energy density would have generally greatly exceeded the dark energy density because such energy, by violating the strong energy condition, decreases far more slowly as the Universe expands than does matter.

This can be seen as follows. For a flat Universe with fraction $\Omega_0$ in matter density, and $\Omega_x$ in radiation at the present time, the age $t_z$ at redshift $z$ is given by

---

[1a] A work which appeared as this was being prepared for submission[23] argues on that the redshift of globular cluster formation can be pushed back to no earlier than z ~7 in extreme models, although a redshift of z=3 is probably favored.

$$H_0 t_{z'} = \int_{z'}^{\infty} \frac{dz}{(1+z)[\Omega_0(1+z)^3 + \Omega_x(1+z)^{3(1+w)}]^{1/2}} \quad (1)$$

where $w = p/\rho$ represents the equation of state for the dark energy (assumed here for simplicity to be constant). For $w < -0.5$, as required in order to produce an accelerating universe, and for $\Omega_0 \sim 0.3$ and $\Omega_x \sim 0.7$ as suggested by observation, the second term is negligible compared to the first for all redshifts greater than about 4.

This relation implies that the age of the Universe at a redshift $z=6$ was greater than approximately 0.8 Gyr, independent of cosmological model. Using this relation, and the estimates given above, we determine, at the 95 % confidence level, a lower limit on the age of the universe of 11 Gyr, and a best-fit age of 13.2 Gyr, based on our best current estimates of the age of the oldest globular clusters in our galaxy.

These limits can be compared with the inferred Hubble age of the universe, given by equation (1) for $z = 0$, for different values of w, and for different values of $H_0$ today. Using the Hubble Key Project estimated range for $H_0 = 70 - 7$ km s$^{-1}$ Mpc$^{-1}$, and recognizing that our estimate of $M_v$-RR produces a change of less than 1 % in $H_0$ compared to the value utilized by the Key Project team, we can determine the range of w allowed by our age estimate. These are shown in Figures 2 and 3, with $H_0 = 70$ and $H_0 = 63$ respectively.

As can be seen, while the size of the allowed range is strongly sensitive to the assumed value of the Hubble constant, our globular cluster age estimate incorporating astrophysical uncertainties now definitively rules out a flat matter-dominated (i.e. w=0) universe at the 95 % confidence level. Interestingly, for the best-fit value of the Hubble constant our result also puts strong limits on the total matter density of the Universe. In order to achieve consistency the mass fraction cannot exceed 38 % of the critical density at the 68% confidence level and 58% of the critical density at the 95% confidence level.

121212

**Conclusions**

Our new estimates for the age of globular clusters significantly increase the lower bound on the age of the universe, and in so doing now provide additional independent evidence that we cannot live in a flat matter-dominated universe. This returns what was once termed as the globular cluster age problem to its former role of providing support for a dark energy component to the universe. Indeed, if we live in a flat universe, the dominant energy density in the universe must reside in empty space. This conclusion bolsters direct observations of the acceleration rate of the universe based on Type 1a supernovae.

While we can now provide further strong evidence in favor of dark energy in the universe, globular cluster dating techniques still have sufficient residual uncertainty so that the constraints on the actual equation of state are not extremely stringent. Nevertheless, it is significant that for a Hubble constant in excess of 70, globular cluster dating provides a new strong constraint on the matter density in a flat universe, independent of the specific equation of state for the dark energy.

1. Chaboyer, B., Kernan, P., Demarque, P., Krauss, L.M. A Lower Limit on the Age of the Universe, *Science,* **271**, 957-960 (1996).

2. Chaboyer, B., Kernan, P., Demarque, P., Krauss, L.M. The Age of Globular Clusters in Light of Hipparos: Resolving the Age Problem?,*Astrophys J.,* **494**, 96-108 (1998)

3. Krauss, L.M. The Age of Globular Clusters, *Physics Reports*, **33,** 333-334 (2000).

4. i.e. see De Bernadis et al, Multiple Peaks in the Angular Power Spectrum of the Cosmic Microwave Background: Significance and Consequences for Cosmology, *Astrophys. J.,* in press (2001)

Table 1: Various Estimates for $M_v(RR)$

| Method | Objects | <Fe/H> | $M_v(RR)$ | $M_v(RR)$ at [Fe/H] = -1.90 | Reference |
|---|---|---|---|---|---|
| Theoretical HB Models | Meta-poor blue HB clusters | -1.90 | 0.35–0.10 | 0.35–0.10 | Demarque et al. 2000[6] |
| Main Sequence Fitting | M13, M68, M92, N6752 | -1.83 | 0.35–0.10 | 0.36–0.10 | Chaboyer 1998[24] |
| White Dwarf Fitting | N6397 | -1.42 | 0.52–0.15 | 0.41–0.15 | Renzini et al. 1996[25] |
| LMC RR Lyr | RR Lyr in the LMC | -1.90 | 0.44–0.10 | 0.44–0.10 | Walker 1992[26] |
| Trigometric Parallax | RR Lyrae | -1.39 | 0.61–0.11 | 0.49–0.11 | Benedict et al. 2001[9] |
| Trigometric Parallax | Field Halo HB Stars | -1.51 | 0.60–0.12 | 0.51–0.12 | Gratton 1998[8] |
| Dynamical | M2, M13, M22, M92 | -1.74 | 0.59–0.15 | 0.55–0.15 | Rees 1996[27] Chaboyer 1998[24] |
| Statistical Parallax | Field RR Lyr | -1.60 | 0.77–0.13 | 0.71–0.13 | Popowski & Gould 1998[28] |

Figure 1: Histogram representing results of Monte Carlo presenting 10,000 fits of predicted isochrones for differing input parameters to observed isochrones to determine the age of the oldest globular clusters.

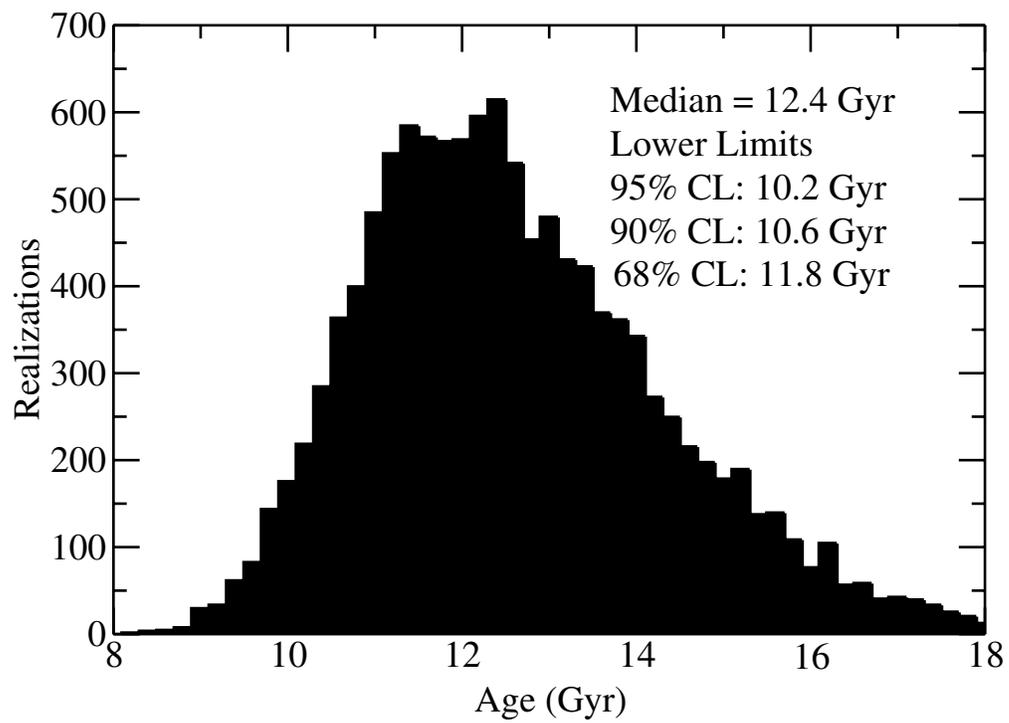

Figure 2: Range of allowed values for the dark-energy equation of state vs. the matter density assuming a flat universe, for the lower limit derived in the text for the age of the Universe, and for $H_0$ =70 km s$^{-1}$Mpc$^{-1}$.

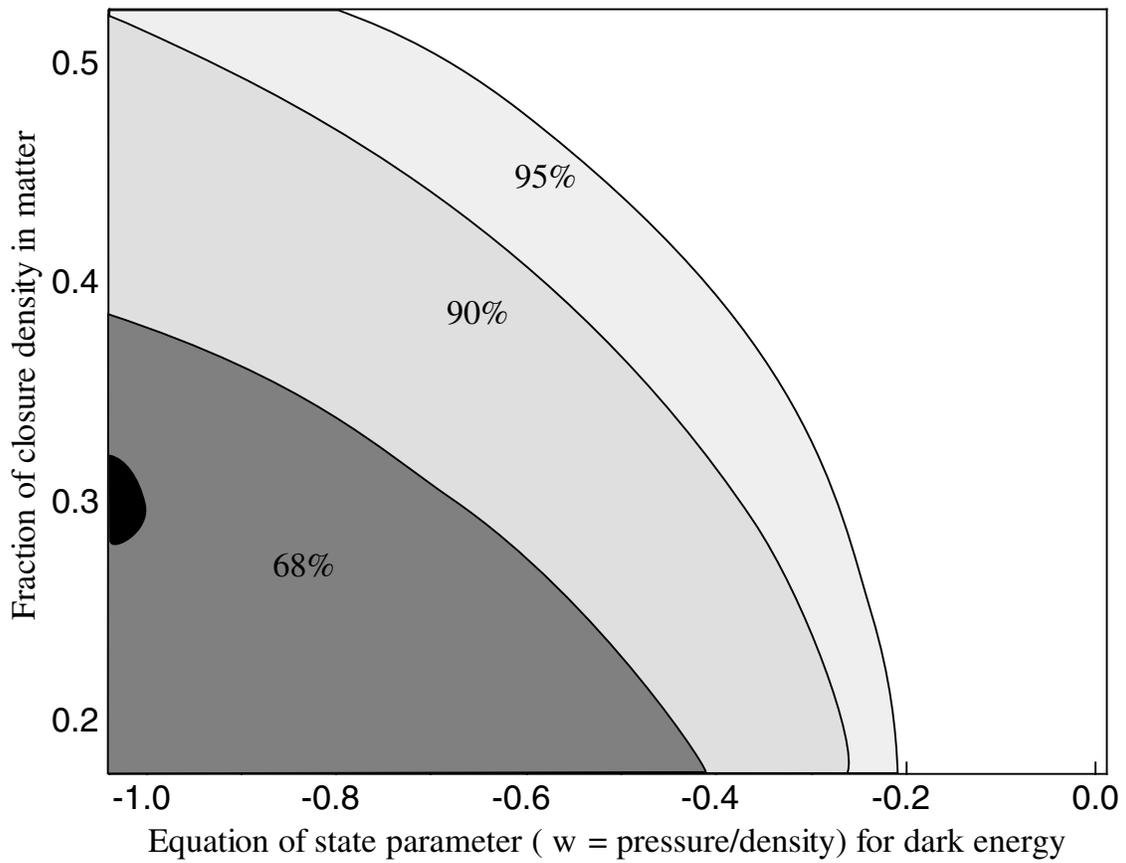

Figure 3: same as Figure 2, for $H_0$ =63 km s$^{-1}$Mpc$^{-1}$.

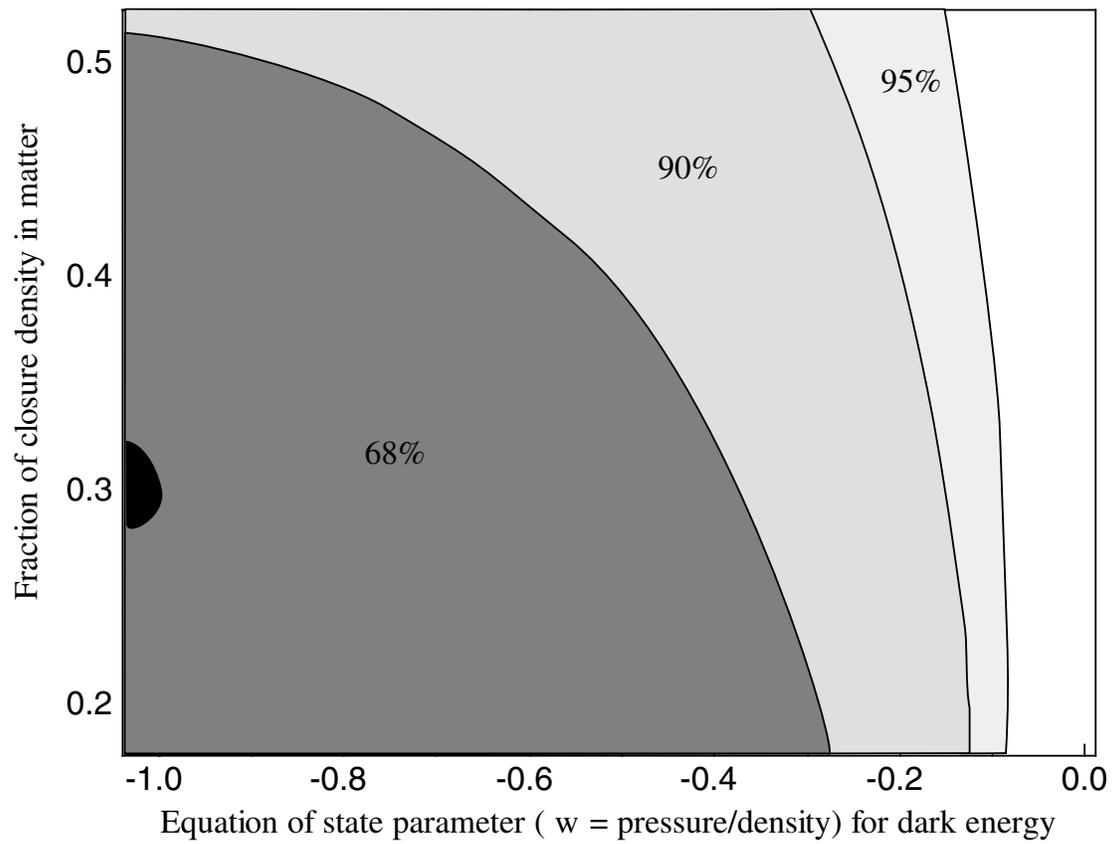